\documentclass[aps,prd,superscriptaddress]{revtex4}
\usepackage{epsfig,epsf}
\usepackage{amsmath}
\usepackage{amsthm}
\usepackage{amsfonts}
\usepackage{amssymb}
\usepackage{dsfont}
\usepackage{multirow}
\usepackage{appendix}
\usepackage{slashed}
\usepackage[active]{srcltx}
\usepackage{psfrag}

\begin{document}

\title{Resonance $X(6600)$}
\date{\today }
\author{S.~S.~Agaev}
\affiliation{Institute for Physical Problems, Baku State University, Az--1148 Baku,
Azerbaijan}
\author{K.~Azizi}
\affiliation{Department of Physics, University of Tehran, North Karegar Avenue, Tehran
14395-547, Iran}
\affiliation{Department of Physics, Dogus University, Dudullu-\"{U}mraniye, 34775
Istanbul, T\"{u}rkiye}
\author{H.~Sundu}
\affiliation{Department of Physics Engineering, Istanbul Medeniyet University, 34700
Istanbul, T\"{u}rkiye}

\begin{abstract}
The resonance $X(6600)$ is explored as the all-charm tetraquark structure
with spin-parities $J^{\mathrm{PC}}=2^{++}$. It is considered in the
diquark-antidiquark picture and modeled as a tensor state $X$ composed of
the axial-vector diquark $cC\gamma_{\mu}c$ and antidiquark $\overline{c}%
\gamma_{\nu}C\overline{c}$ with $C$ being the charge conjugation matrix. The
mass and decay width of $X$ are evaluated in the framework of QCD sum rule
(SR) methods. The two-point SR approach is applied to find its spectroscopic
parameters, while three-point SRs used to calculate partial widths of
different decay channels of $X$. We study its leading decays $X \to J/\psi
J/\psi$, $X \to \eta_{c}\eta_{c}$ and $\chi _{c1}(1P)\eta _{c}$ in which all
four $c$-quarks constitute final-state mesons. We consider also the
subleading channels $X\to D_{(s)}^{(\ast )+}D_{(s)}^{(\ast )-}$ and $%
D_{(s)}^{(\ast )0}\overline{D}_{(s)}^{(\ast )0}$ generated by annihilation
of $\overline{c}c$ quarks in the tetraquark. Comparison of the mass $m=(6609
\pm 50)~ \mathrm{MeV}$ and width $\Gamma[X]=(165 \pm 23)~ \mathrm{MeV}$ of
the tensor diquark-antidiquark state $X$ with experimental data allows us to
interpret it as an essential component of the resonance $X(6600)$. We also
provide a lower limit for the mass of the first radial excitation of $X$.
\end{abstract}

\maketitle


\section{Introduction}

\label{sec:Intro} 

The experimental discoveries of four heavy $X(6200)$, $X(6600)$, $X(6900)$,
and $X(7300)/X(7100)$ structures made during last years by LHCb-ATLAS-CMS
collaborations had important impact on physics of exotic multiquark mesons
and, in particular, on the spectroscopy of fully-charmed tetraquarks \cite%
{LHCb:2020bwg,ATLAS:2023bft,CMS:2023owd,CMS:2026tiu}. They were observed in
the di-$J/\psi $ and $J/\psi \psi (2S)$ mass distributions and are
presumably resonances built of $cc\overline{c}\overline{c}$ quarks.

These discoveries stimulated new detailed investigations of all-heavy exotic
mesons \cite{Bedolla:2019zwg,Zhang:2020xtb,Wang:2020ols,Wang:2020dlo,
Albuquerque:2020hio,Yang:2020wkh,Cordillo:2020sgc,Dong:2020nwy,Dong:2021lkh,Liang:2021fzr,Wang:2021kfv,Deng:2020iqw}. Theoretical interpretations of these states, calculations of their masses
and other parameters using numerous models and methods, comparison of
obtained results with experimental data led to gaining valuable knowledge
about all-charm tetraquarks. In these studies the $X$ resonances were
treated mainly as diquark-antidiquark and hadronic molecule states with
different spin-parities.

Predictions for parameters of these states provide information necessary for
comparison with available data. At the same time, they depend on models
applied to study all-charm tetraquarks. Thus, the mass of a ground-state
scalar tetraquark $cc\overline{c}\overline{c}$ was estimated in the
relativized diquark model as $m=5883~\mathrm{GeV}$ \cite{Bedolla:2019zwg}.
Interesting results were obtained by means of the sum rule (SR) method \cite%
{Zhang:2020xtb,Wang:2020ols,Wang:2020dlo, Albuquerque:2020hio,Yang:2020wkh}.
For instance, the mass of the scalar tetraquark $cc\overline{c}\overline{c}$
was predicted within limits $6.44-6.47~\mathrm{GeV}$ \cite{Zhang:2020xtb}.
In Ref.\ \cite{Albuquerque:2020hio} fully-heavy states were analyzed in the
diquark-antidiquark and molecular frameworks, and the authors interpreted $%
X(6900)$ as a hadronic molecule $\chi _{c0}\chi _{c0}$ or/and a tetraquark
composed of pseudoscalar components.

It should be noted that alternative explanations of the observed structures
were suggested as well. Indeed, new $X$ states were considered in Ref.\ \cite%
{Dong:2020nwy} within a coupled-channel approach. It was argued that in the
di-$J/\psi $ system exists a near-threshold state $X(6200)$ with
spin-parities $0^{++}$ or $2^{++}$. The resonance $X(6900)$ may also be
generated as a pole structure due to coupled-channel effects \cite%
{Liang:2021fzr}. The authors of this article predicted existence of a bound
state $X(6200)$, and broad and narrow resonances $X(6680)$ and $X(7200)$,
respectively.

Various assumptions about nature of the $X$ resonances were also made in
Refs.\ \cite%
{Wang:2022xja,Faustov:2022mvs,Niu:2022vqp,Dong:2022sef,Yu:2022lak,An:2022qpt,Kuang:2023vac,Liu:2020eha,Malekhosseini:2025hyx}. Thus, possible assignments of $X(6600)$, $X(6900)$ and $X(7300)$
resonances in the diquark-antidiquark model with the quantum numbers $J^{%
\mathrm{PC}}=0^{++}$ or $1^{+-}$ were explored in Ref.\ \cite{Wang:2022xja}.
In this work the author applied the SR method and Regge trajectory
techniques. In the framework of the relativistic quark model, the resonance $%
X(6200)$ was considered as a ground-level scalar tetraquark $J^{\mathrm{PC}%
}=0^{++}$, whereas $X(6600)$ identified with either $1S$ tensor state $J^{%
\mathrm{PC}}=2^{++}$ or a radial excitation ($2S$) of the scalar structure.
The $X(6900)$ state maybe a $2S$ excited tensor tetraquark or $1D$
scalar/tensor exotic mesons \cite{Faustov:2022mvs}. In the another context,
similar assignments were also proposed in Ref.\ \cite{Dong:2022sef} .

The $X$ resonances were studied in our works \cite%
{Agaev:2023wua,Agaev:2023ruu,Agaev:2023gaq,Agaev:2023rpj} as well, where we
evaluated parameters of various all-charmed scalar structures. Analyses were
done using SR method and diquark-antidiquark and hadronic molecule models.
It was found that the resonance $X(6200)$ is presumably the $\eta _{c}\eta
_{c}$ \cite{Agaev:2023ruu} molecule. The structure $X(6600)$ was considered
in the diquark-antidiquark picture as a state built of axial-vector
components \cite{Agaev:2023wua}. Properties of resonance $X(6900)$, i.e.,
its mass and full width agree with data provided one models it as a
tetraquark composed of pseudoscalar diquarks and/or as a hadronic molecule $%
\chi _{c0}\chi _{c0}$ \cite{Agaev:2023ruu,Agaev:2023gaq}. Therefore, it was
interpreted as an admixture of the diquark-antidiquark and molecule-type
states. The last resonance from this list $X(7300)$ was treated as a
superposion of the hadronic molecule $\chi _{c1}\chi _{c1}$ and first radial
excitation of $X(6600)$ \cite{Agaev:2023rpj}.

In general, the $X$ resonances were mainly explored as scalar particles with
spin-parities $J^{\mathrm{PC}}=0^{++}$. But recently the CMS collaboration
reported on the first measurements of $X$ resonances' quantum numbers \cite%
{CMS:2025fpt}. The parity $\mathrm{P}$ and charge conjugation $\mathrm{C}$
symmetries of these tetraquarks were found to be $+1$. Their spin $J$ was
determined to be consistent with $J=2$, whereas $J=0$ and $J=1$ options were
excluded at $95\%$ and $99\%$ confidence levels, respectively. The
collaboration also made suggestions about diquark-antidiquark nature of the
resonance $X(6600)$ and considered states $X(6900)$ and $X(7100)$ as its
radial excitations.

In present article, in light of these experimental measurements, we explore
the tensor diquark-antidiquark state $X=cc\overline{c}\overline{c}$ with
internal organization $C\gamma _{\mu }\otimes \gamma _{\nu }C+C\gamma _{\nu
}\otimes \gamma _{\mu }C$. We calculate its mass and current coupling by
employing QCD two-point SR method \cite{Shifman:1978bx,Shifman:1978by},
which is one of effective tools to determine parameters of numerous hadrons.
It was originally invented to investigate conventional mesons and baryons,
but successfully applied for analyses of multiquark, gluon-quark, and pure
gluon systems as well \cite{Albuquerque:2018jkn,Agaev:2020zad,Wang:2025sic}.

Information obtained during this analysis permits us to find also its
kinematically allowed decay channels. It turns out that the tensor
tetraquark $X$ is unstable against strong decays to ordinary meson pairs $%
J/\psi J/\psi $,$\ \eta _{c}\eta _{c}$, and $\chi _{c1}(1P)\eta _{c}$. These
processes are leading decay channels of $X$ because all initial $c$ quarks
(antiquarks) appear in the final-state mesons. But there is another
mechanism for transformation of the tetraquark $X$ through strong
interactions. Namely,$\ $an annihilation of $c\overline{c}$ pairs from $X$
to light quarks $q\overline{q}$, $s\overline{s}$ followed by production of $%
D_{(s)}^{(\ast )+}D_{(s)}^{(\ast )-}$ and $D_{(s)}^{(\ast )0}\overline{D}%
_{(s)}^{(\ast )0}$ mesons generates second type of channels which we refer
to as subleading decays of the tetraquark $X$ \cite%
{Becchi:2020mjz,Becchi:2020uvq,Agaev:2023ara}.

The decays of $X$ are explored by employing the technical tools of the
three-point sum rule approach. It allows one to evaluate strong couplings $%
g_{i}$ at the tetraquark-meson-meson vertices necessary to calculate partial
widths of processes under analysis.

This paper contains five sections: The mass and current coupling of the
tensor tetraquark $X$ are calculated in Sec.\ \ref{sec:Mass}. The leading
decay channels $X\rightarrow J/\psi J/\psi $, $\eta _{c}\eta _{c}$, and $%
\chi _{c1}(1P)\eta _{c}$ are considered in Sec.\ \ref{sec:Widths1}. The
section \ref{sec:Widths2} is devoted to analyses of subleading modes of $X$.
In this section we compute partial widths of six channels. Here, we evaluate
the full width of $X$ as well. The last Sec. \ref{sec:Conc} is reserved for
discussions and final notes.  The Appendix contains explicit expressions of the correlation function corresponding to the mass sum rule.


\section{Spectroscopic parameters of $X$}

\label{sec:Mass} 

The mass $m$ and current coupling (pole residue) $\Lambda $ of the tensor
tetraquark $X$ can be evaluated using the relevant sum rules for these
parameters. The required sum rules are obtained from exploration of the
two-point correlation function 
\begin{equation}
\Pi _{\mu \nu \alpha \beta }(p)=i\int d^{4}xe^{ipx}\langle 0|\mathcal{T}%
\{I_{\mu \nu }(x)I_{\alpha \beta }^{\dag }(0)\}|0\rangle ,  \label{eq:CF1}
\end{equation}%
where $I_{\mu \nu }(x)$ is the interpolating current for the tensor state $X$%
.

We model $X$ as a structure built of axial-vector diquark and antidiquark
components, therefore $I_{\mu \nu }(x)$ has the following form 
\begin{equation}
I_{\mu \nu }(x)=c_{a}^{T}(x)C\gamma _{\mu }c_{b}(x)\overline{c}_{a}(x)\gamma
_{\nu }C\overline{c}_{b}^{T}(x)+c_{a}^{T}(x)C\gamma _{\nu }c_{b}(x)\overline{%
c}_{a}(x)\gamma _{\mu }C\overline{c}_{b}^{T}(x).
\end{equation}%
Here, $c(x)$ is $c$-quark field with $a$ and $b$ being the color indices,
and $C$ is the charge conjugation matrix. The current $I_{\mu \nu }(x)$
describes the particle with the spin-parities $J^{\mathrm{PC}}=2^{++}$.

To find SRs for the parameters $m$ and $\Lambda $, one should compute the
correlation function $\Pi _{\mu \nu \alpha \beta }(p)$ using two approaches.
In the first one, the correlator is expressed in terms of physical
parameters of the tetraquark which leads to physical component $\Pi _{\mu
\nu \alpha \beta }^{\mathrm{Phys}}(p)$ of SRs. To this end, we insert into $%
\Pi _{\mu \nu \alpha \beta }(p)$ a complete set of intermediate states, and
carry out the integration over $x$. As a result, we get 
\begin{equation}
\Pi _{\mu \nu \alpha \beta }^{\mathrm{Phys}}(p)=\frac{\langle 0|I_{\mu \nu
}|X(p,\epsilon )\rangle \langle X(p,\epsilon )|I_{\alpha \beta }^{\dag
}|0\rangle }{m^{2}-p^{2}}+\cdots .
\end{equation}%
where $\epsilon =\epsilon _{\mu \nu }(p)$ is the polarization tensor of the
tetraquark $X$. Above, the term that corresponds to ground-state particle $X$
is presented explicitly, whereas contributions of higher resonances and
continuum states are shown by the ellipses.

The formula for $\Pi _{\mu \nu \alpha \beta }^{\mathrm{Phys}}(p)$ can be
simplified by employing the matrix element 
\begin{equation}
\langle 0|I_{\mu \nu }|X(p,\epsilon (p))\rangle =\Lambda \epsilon _{\mu \nu
}(p).  \label{eq:ME1}
\end{equation}%
Having used Eq. (\ref{eq:ME1}) in the correlator $\Pi _{\mu \nu \alpha \beta
}^{\mathrm{Phys}}(p)$ and performed required operations, one finds 
\begin{equation}
\Pi _{\mu \nu \alpha \beta }^{\mathrm{Phys}}(p)=\frac{\Lambda ^{2}}{%
m^{2}-p^{2}}\left\{ \frac{1}{2}\left( g_{\mu \alpha }g_{\nu \beta }+g_{\mu
\beta }g_{\nu \alpha }\right) +\text{ other components}\right\} +\cdots .
\label{eq:Phys2a}
\end{equation}%
The function $\Pi _{\mu \nu \alpha \beta }^{\mathrm{Phys}}(p)$ is composed
of contributions with different Lorentz structures. The term $\sim (g_{\mu
\alpha }g_{\nu \beta }+g_{\mu \beta }g_{\nu \alpha })$ arises from a spin-$2$
particle. Therefore, it is convenient to employ this term and corresponding
invariant amplitude $\Pi ^{\mathrm{Phys}}(p^{2})$ in our analysis.

To determine QCD side of the sum rules $\Pi _{\mu \nu \alpha \beta }^{%
\mathrm{OPE}}(p)$ we insert $I_{\mu \nu }(x)$ into $\Pi _{\mu \nu \alpha
\beta }(p)$ and contract quark fields. We get 
\begin{eqnarray}
&&\Pi _{\mu \nu \alpha \beta }^{\mathrm{OPE}}(p)=i\int d^{4}xe^{ipx}\left\{
\left\{ \mathrm{Tr}\left[ \gamma _{\nu }\widetilde{S}_{c}^{b^{\prime
}b}(-x)\gamma _{\beta }S_{c}^{a^{\prime }a}(-x)\right] -\mathrm{Tr}\left[
\gamma _{\nu }\widetilde{S}_{c}^{a^{\prime }b}(-x)\gamma _{\beta
}S_{c}^{b^{\prime }a}(-x)\right] \right\} \right.   \notag \\
&&\times \left\{ \mathrm{Tr}\left[ \gamma _{\alpha }\widetilde{S}%
_{c}^{a^{\prime }a}(x)\gamma _{\mu }S_{c}^{bb^{\prime }}(x)\right] -\mathrm{%
Tr}\left[ \gamma _{\alpha }\widetilde{S}_{c}^{ba^{\prime }}(x)\gamma _{\mu
}S_{c}^{ab^{\prime }}(x)\right] \right\} +\left( \mu \leftrightarrow \nu
\right) +\left( \alpha \leftrightarrow \beta \right)   \notag \\
&&\left. +\left( \mu \leftrightarrow \nu ,\alpha \leftrightarrow \beta
\right) \right\} ,  \label{eq:QCD1a}
\end{eqnarray}%
where $S_{c}(x)$ is $c-$quark propagator \cite{Agaev:2020zad}, and $%
\widetilde{S}_{c}(x)$ is defined by the expression%
\begin{equation}
\widetilde{S}_{c}(x)=CS_{c}^{T}(x)C.
\end{equation}

The correlator $\Pi _{\mu \nu \alpha \beta }^{\mathrm{OPE}}(p)$ has to be
computed using operator product expansion ($\mathrm{OPE}$) with some
accuracy. Having extracted in $\Pi _{\mu \nu \alpha \beta }^{\mathrm{OPE}%
}(p) $ a contribution proportional to $(g_{\mu \alpha }g_{\nu \beta }+g_{\mu
\beta }g_{\nu \alpha })$ and labeled by $\Pi ^{\mathrm{OPE}}(p^{2})$ the
relevant amplitude, one can determine the required SRs. To this end, one
equates the amplitudes $\Pi ^{\mathrm{Phys}}(p^{2})$ and $\Pi ^{\mathrm{OPE}%
}(p^{2})$ and performs standard operations of SR method. In other words, one
applies the Borel transformation which suppresses contributions of higher
resonances and continuum states. Afterwards, using the quark-hadron duality
assumption, one subtracts these contributions from QCD side of the SR
equality. After these manipulations $\Pi ^{\mathrm{OPE}}(p^{2})$ becomes
equal to $\Pi (M^{2},s_{0})$ and depends on the Borel and continuum
subtraction parameters $M^{2}$ and $s_{0}$. The SRs for the mass $m$ and
current coupling $\Lambda $ read 
\begin{equation}
m^{2}=\frac{\Pi ^{\prime }(M^{2},s_{0})}{\Pi (M^{2},s_{0})},  \label{eq:Mass}
\end{equation}%
and 
\begin{equation}
\Lambda ^{2}=e^{m^{2}/M^{2}}\Pi (M^{2},s_{0}),  \label{eq:Coupl}
\end{equation}%
where $\Pi ^{\prime }(M^{2},s_{0})=d\Pi (M^{2},s_{0})/d(-1/M^{2})$.

The transformed amplitude $\Pi (M^{2},s_{0})$ is given by the formula%
\begin{equation}
\Pi (M^{2},s_{0})=\int_{16m_{c}^{2}}^{s_{0}}ds\rho ^{\mathrm{OPE}%
}(s)e^{-s/M^{2}}+\Pi (M^{2}).  \label{eq:CorrF}
\end{equation}%
In Eq.\ (\ref{eq:CorrF}) $\rho ^{\mathrm{OPE}}(s)$ is the two-point spectral
density which amounts to the imaginary part of the amplitude $\Pi ^{\mathrm{%
OPE}}(p^{2})$. The contribution $\Pi (M^{2})$ is obtained directly from $\Pi
^{\mathrm{OPE}}(p^{2})$ and contains terms absent in $\rho ^{\mathrm{OPE}}(s)
$. We compute $\Pi (M^{2},s_{0})$ by taking into account dimension-$4$ terms 
$\sim \langle \alpha _{s}G^{2}/\pi \rangle $. Note that in fully-heavy
systems in absence of light quark and mixed condensates the triple-gluon
condensate $\langle g_{s}^{3}G^{3}\rangle $ is the next nonperturbative
contribution. But in fully-heavy systems dimension-four effects are already
very small therefore terms $\sim \langle g_{s}^{3}G^{3}\rangle $ can be
safely neglected. This conclusion was confirmed once more by explicit
calculations in Ref. \cite{Agaev:2025did}.

Then $\rho ^{\mathrm{OPE}}(s)$ is equal to $\rho ^{\mathrm{OPE}}(s)=\rho ^{%
\mathrm{pert.}}(s)+\rho ^{\mathrm{Dim4}}(s)$, whereas $\Pi (M^{2})$
contributes starting from $\mathrm{Dim4}$ term. In Appendix we provide
analytical expressions of functions $\rho ^{\mathrm{pert.}}(s)$ and $\Pi ^{%
\mathrm{Dim4}}(M^{2})$ and refrain from presenting $\rho ^{\mathrm{Dim4}}(s)$
which is rather cumbersome.

For numerical analysis we should determine parameters in the relevant SRs.
For the mass $m_{c}$ of $c$ quark and gluon condensate $\langle \alpha
_{s}G^{2}/\pi \rangle $ we employ%
\begin{equation}
m_{c}=(1.2730\pm 0.0046)~\mathrm{GeV},\ \langle \alpha _{s}G^{2}/\pi \rangle
=(0.012\pm 0.004)~\mathrm{GeV}^{4},
\end{equation}%
which are universal entries.

The parameters $M^{2}$ and $s_{0}$ depend on the process under investigation
and have to satisfy usual restrictions of the sum rule analysis. These
constraints imply dominance of the pole contribution ($\mathrm{PC}$) in
physical quantities, convergence of $\mathrm{OPE}$ and minimal dependence of 
$m$ and $\Lambda $ on parameters $M^{2}$ and $s_{0}$: These constants are
important for the credibility of SR results. Therefore, we require
fulfillment of $\mathrm{PC}\geq 0.5$, where 
\begin{equation}
\mathrm{PC}=\frac{\Pi (M^{2},s_{0})}{\Pi (M^{2},\infty )}.  \label{eq:PC}
\end{equation}

Because $\Pi (M^{2},s_{0})$ contains the perturbative and dimension-$4$
contribution $\Pi ^{\mathrm{Dim4}}(M^{2},s_{0})$, to ensure convergence of $%
\mathrm{OPE}$, we impose the constraint $|\Pi ^{\mathrm{Dim4}%
}(M^{2},s_{0})|\leq 0.05|\Pi (M^{2},s_{0})|$. Note that these two
restrictions allow us to find the maximal and minimal values of $M^{2}$,
respectively.

\begin{figure}[h]
\includegraphics[width=8.5cm]{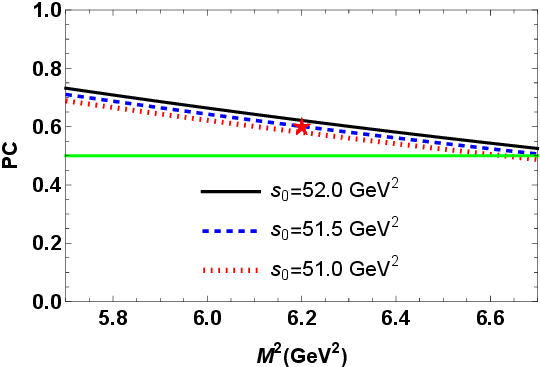}
\caption{$\mathrm{PC}$ as a function of $M^{2}$ at some $s_{0}$. The
horizontal line shows the border $\mathrm{PC}=0.5$. The star fixes the point 
$M^{2}=6.2~\mathrm{GeV}^{2},s_{0}=51.5~\mathrm{GeV}^{2}$. }
\label{fig:PC}
\end{figure}

Our computations demonstrate that regions for $M^{2}$ and $s_{0}$ 
\begin{equation}
M^{2}\in \lbrack 5.7,6.7]~\mathrm{GeV}^{2},\ s_{0}\in \lbrack 51,52]~\mathrm{%
GeV}^{2}  \label{eq:Wind1}
\end{equation}%
satisfy all necessary constraints. In fact, the pole contribution on the
average in $s_{0}$ is $\mathrm{PC}\approx 0.50$ and $\mathrm{PC}\approx 0.71$
at $6.7~\mathrm{GeV}^{2}$ and $5.7~\mathrm{GeV}^{2}$, respectively. The term 
$|\Pi ^{\mathrm{Dim4}}(M^{2},s_{0})|$ at $M^{2}=5.7~\mathrm{GeV}^{2}$ is
around $0.7\%$ of the amplitude $\Pi (M^{2},s_{0})$. In Fig.\ \ref{fig:PC}
we show $\mathrm{PC}$ as a function of the Borel parameter $M^{2}$ in which
almost all lines exceed limit $\mathrm{PC}=0.5$. 

\begin{widetext}

\begin{figure}[htbp]
\begin{center}
\includegraphics[totalheight=6cm,width=8cm]{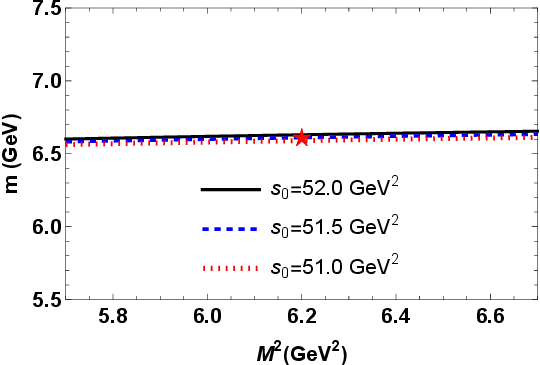}
\includegraphics[totalheight=6cm,width=8cm]{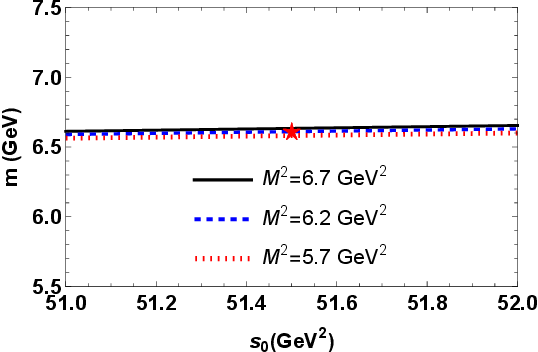}
\end{center}
\caption{Dependence of the mass $m$ on the parameters $M^{2}$ (left), and $s_0$ (right).}
\label{fig:Mass}
\end{figure}

\end{widetext}

We evaluate $m$ and $\Lambda $ in the regions Eq.\ (\ref{eq:Wind1}) and
determine their average values 
\begin{equation}
m=(6609\pm 50)~\mathrm{MeV},\ \Lambda =(4.25\pm 0.28)\times 10^{-1}~\mathrm{%
GeV}^{5}.  \label{eq:Result1a}
\end{equation}%
The predictions in Eq.\ (\ref{eq:Result1a}) are equivalent to SR results at
the point $M^{2}=6.2~\mathrm{GeV}^{2}$ and $s_{0}=51.5~\mathrm{GeV}^{2}$,
where $\mathrm{PC}\approx 0.6$. This guaranties prevalence of $\mathrm{PC}$
in the quantities $m$ and $\Lambda $. Ambiguities in Eq.\ (\ref{eq:Result1a}%
) are connected with those in $M^{2}$ and $s_{0}$: The quark mass $m_{c}$
and condensate $\langle \alpha _{s}G^{2}/\pi \rangle $ almost do not
generate sizeable errors. Uncertainties of the mass $m$ are equal to $\pm
0.8\%$ of the central value, whereas for $\Lambda $ they amount to $\pm 6.6\%
$. All these theoretical errors are inside of usual limits of SR analysis
proving validity of the obtained results. Dependencies of the mass $m$ on
the Borel and continuum subtraction parameters are depicted in Fig.\ \ref%
{fig:Mass}.


\section{Leading decay channels of the tetraquark $X$}

\label{sec:Widths1}


Result for the mass $m$ of the tensor tetraquark $X$ and its quantum numbers 
$J^{\mathrm{PC}}=2^{++}$ allow us to find its leading (dominant) decay
channels. One can see that decays to pairs of the vector $J/\psi $ and
pseudoscalar $\eta _{c}$ mesons, as well as to $\chi _{c1}(1P)\eta _{c}$
mesons are the kinematically permitted modes of the tetraquark $X$. In fact,
two-meson thresholds for these processes $6194~$\textrm{M}$\mathrm{eV}$, $%
5969~\mathrm{MeV}$ and $6495~\mathrm{MeV}$ are below the mass of $X$. Even
in lower limit of the mass $m=6559~\mathrm{MeV}$ the tetraquark $X$ easily
decays through these channels.


\subsection{ $X\rightarrow J/\protect\psi J/\protect\psi $}


Here, we consider the mode $X\rightarrow J/\psi J/\psi $ and calculate its
partial width. For these purposes, we first evaluate the strong coupling $%
g_{1}$ that describes the interaction of particles at the vertex $XJ/\psi
J/\psi $. In its turn, $g_{1}$ is equal to the form factor $g_{1}(q^{2})$ at
the mass shell $q^{2}=m_{J/\psi }^{2}$. The latter can be obtained from
exploration of the three-point correlator 
\begin{equation}
\Pi _{\mu \nu \alpha \beta }(p,p^{\prime })=i^{2}\int
d^{4}xd^{4}ye^{ip^{\prime }y}e^{-ipx}\langle 0|\mathcal{T}\{I_{\mu }^{J/\psi
}(y)I_{\nu }^{J/\psi }(0)I_{\alpha \beta }^{\dagger }(x)\}|0\rangle ,
\label{eq:CF1aa}
\end{equation}%
which is necessary to derive SR for $g_{1}(q^{2})$. In Eq.\ (\ref{eq:CF1aa}) 
$I_{\mu }^{J/\psi }(x)$ is the current which interpolate the vector
charmonium $J/\psi $ $\ $%
\begin{equation}
I_{\mu }^{J/\psi }(x)=\overline{c}_{i}(x)\gamma _{\mu }c_{i}(x),
\end{equation}%
with $i$ being the color index.

The correlation function $\Pi _{\mu \nu \alpha \beta }(p,p^{\prime })$ once
expressed in parameters of the particles $X$ and $J/\psi $ gives the
phenomenological side $\Pi _{\mu \nu \alpha \beta }^{\mathrm{Phys}%
}(p,p^{\prime })$ of SR. Having taken into account contribution of the
ground-state particles, we get%
\begin{eqnarray}
&&\Pi _{\mu \nu \alpha \beta }^{\mathrm{Phys}}(p,p^{\prime })=\frac{\langle
0|I_{\mu }^{J/\psi }|J/\psi (p^{\prime },\varepsilon _{1})\rangle }{%
p^{\prime 2}-m_{J/\psi }^{2}}\frac{\langle 0|I_{\nu }^{J/\psi }|J/\psi
(q,\varepsilon _{2})\rangle }{q^{2}-m_{J/\psi }^{2}}\langle J/\psi
(p^{\prime },\varepsilon _{1})J/\psi (q,\varepsilon _{2})|X(p,\epsilon
)\rangle  \notag \\
&&\times \frac{\langle X(p,\varepsilon )|I_{\alpha \beta }^{\dagger
}|0\rangle }{p^{2}-m^{2}}+\cdots .  \label{eq:TP1}
\end{eqnarray}%
Above, $m_{J/\psi }=(3096.900\pm 0.006)~\mathrm{MeV}$ is the mass of the
meson $J/\psi $ \cite{PDG:2024}, whereas $\varepsilon _{1}$ and $\varepsilon
_{2}$ are the polarization vectors of $J/\psi $ mesons.

Equation (\ref{eq:TP1}) can be recast into a convenient form. To this end,
we employ the matrix elements%
\begin{equation}
\langle 0|I_{\mu }^{J/\psi }|J/\psi (p^{\prime },\varepsilon _{1})\rangle
=f_{J/\psi }m_{J/\psi }\varepsilon _{1\mu }(p^{\prime }),\ \langle 0|I_{\nu
}^{J/\psi }|J/\psi (q,\varepsilon _{2})\rangle =f_{J/\psi }m_{J/\psi
}\varepsilon _{2\nu }(q),  \label{eq:C2}
\end{equation}%
where $f_{J/\psi }=(411\pm 7)~\mathrm{MeV}$ is the decay constant of $J/\psi 
$ \cite{Lakhina:2006vg}.

Detailed analysis demonstrates that the tensor-vector-vector vertex $\langle
J/\psi (p^{\prime },\varepsilon _{1})J/\psi (q,\varepsilon
_{2})|X(p,\epsilon )\rangle $ can be expressing using the momenta and
polarization vectors of particles $X$ and $J/\psi $ in the form \cite%
{Agaev:2024pil} 
\begin{eqnarray}
&&\langle J/\psi (p^{\prime },\varepsilon _{1})J/\psi (q,\varepsilon
_{2})|X(p,\epsilon )\rangle =g_{1}(q^{2})\epsilon _{\tau \rho }\left[
(\varepsilon _{1}^{\ast }\cdot q)\varepsilon _{2}^{\tau \ast }p^{\prime \rho
}+(\varepsilon _{2}^{\ast }\cdot p^{\prime })\varepsilon _{1}^{\ast \tau
}q^{\rho }-(p^{\prime }\cdot q)\varepsilon _{1}^{\tau \ast }\varepsilon
_{2}^{\rho \ast }\right.  \notag \\
&&\left. -(\varepsilon _{1}^{\ast }\cdot \varepsilon _{2}^{\ast })p^{\prime
\tau }q^{\rho }\right] .  \label{eq:TVV}
\end{eqnarray}%
Then, for $\Pi _{\mu \nu \alpha \beta }^{\mathrm{Phys}}(p,p^{\prime })$ we
get 
\begin{eqnarray}
&&\Pi _{\mu \nu \alpha \beta }^{\mathrm{Phys}}(p,p^{\prime })=g_{1}(q^{2})%
\frac{\Lambda f_{J/\psi }^{2}m_{J/\psi }^{2}}{\left( p^{2}-m^{2}\right)
(p^{\prime 2}-m_{J/\psi }^{2})}\frac{1}{(q^{2}-m_{J/\psi }^{2})}\left[
p_{\beta }^{\prime }p_{\alpha }^{\prime }g_{\mu \nu }+\frac{1}{2}p_{\mu
}p_{\alpha }^{\prime }g_{\beta \nu }\right.  \notag \\
&&\left. +\frac{1}{2m^{2}}p_{\beta }p_{\nu }p_{\mu }^{\prime }p_{\alpha
}^{\prime }+\text{ other terms}\right] +\cdots .
\end{eqnarray}

The correlation function $\Pi _{\mu \nu \alpha \beta }^{\mathrm{OPE}%
}(p,p^{\prime })$ is given by the formula 
\begin{eqnarray}
&&\Pi _{\mu \nu \alpha \beta }^{\mathrm{OPE}}(p,p^{\prime })=2i^{2}\int
d^{4}xd^{4}ye^{ip^{\prime }y}e^{-ipx}\left\{ \mathrm{Tr}\left[ \gamma _{\mu
}S_{c}^{ia}(y-x)\gamma _{\alpha }\widetilde{S}_{c}^{jb}(-x)\gamma _{\nu }%
\widetilde{S}_{c}^{aj}(x)\gamma _{\beta }S_{c}^{bi}(x-y)\right] \right. 
\notag \\
&&\left. -\mathrm{Tr}\left[ \gamma _{\mu }S_{c}^{ia}(y-x)\gamma _{\alpha }%
\widetilde{S}_{c}^{jb}(-x)\gamma _{\nu }\widetilde{S}_{c}^{bj}(x)\gamma
_{\beta }S_{c}^{ai}(x)\right] \right\} .  \label{eq:CF3a}
\end{eqnarray}%
We use the amplitudes $\Pi _{1}^{\mathrm{Phys}}(p^{2},p^{\prime 2},q^{2})$
and $\Pi _{1}^{\mathrm{OPE}}(p^{2},p^{\prime 2},q^{2})$ that correspond to
terms $\sim p_{\beta }p_{\nu }p_{\mu }^{\prime }p_{\alpha }^{\prime }$ in
these correlators, and derive SR for $g_{1}(q^{2})$. Usual manipulations
yield 
\begin{equation}
g_{1}(q^{2})=\frac{2m^{2}(q^{2}-m_{J/\psi }^{2})}{\Lambda f_{J/\psi
}^{2}m_{J/\psi }^{2}}e^{m^{2}/M_{1}^{2}}e^{m_{J/\psi }^{2}/M_{2}^{2}}\Pi
_{1}(\mathbf{M}^{2},\mathbf{s}_{0},q^{2}).  \label{eq:SRG}
\end{equation}%
In Eq.\ (\ref{eq:SRG}), $\Pi _{1}(\mathbf{M}^{2},\mathbf{s}_{0},q^{2})$ is
the function $\Pi _{1}^{\mathrm{OPE}}(p^{2},p^{\prime 2},q^{2})$ after the
Borel transformations and continuum subtractions. It depends on the
parameters $\mathbf{M}^{2}=(M_{1}^{2},M_{2}^{2})$ and $\mathbf{s}%
_{0}=(s_{0},s_{0}^{\prime })$. The pair $(M_{1}^{2},s_{0})$ corresponds to $%
X $ channel, while $(M_{2}^{2},s_{0}^{\prime })$ is related to $J/\psi $
channel. The function $\Pi _{1}(\mathbf{M}^{2},\mathbf{s}_{0},q^{2})$ is
determined as%
\begin{equation}
\Pi _{1}(\mathbf{M}^{2},\mathbf{s}_{0},q^{2})=\int_{16m_{c}^{2}}^{s_{0}}ds%
\int_{4m_{c}^{2}}^{s_{0}^{\prime }}ds^{\prime }\rho _{1}(s,s^{\prime
},q^{2})e^{-s/M_{1}^{2}-s^{\prime }/M_{2}^{2}}.  \label{eq:CorrF1}
\end{equation}

Constraints imposed on $\mathbf{M}^{2}$ and $\mathbf{s}_{0}$ are standard
for the sum rule investigations and have been detailed in the previous
section. Our calculations prove that regions Eq.\ (\ref{eq:Wind1}) for the
parameters $(M_{1}^{2},s_{0})$ and 
\begin{equation}
M_{2}^{2}\in \lbrack 4,5]~\mathrm{GeV}^{2},\ s_{0}^{\prime }\in \lbrack
12,13]~\mathrm{GeV}^{2}.  \label{eq:Wind3}
\end{equation}%
for $(M_{2}^{2},s_{0}^{\prime })$ meet these conditions.

\begin{widetext}

\begin{figure}[h!]
\begin{center}
\includegraphics[totalheight=6cm,width=8cm]{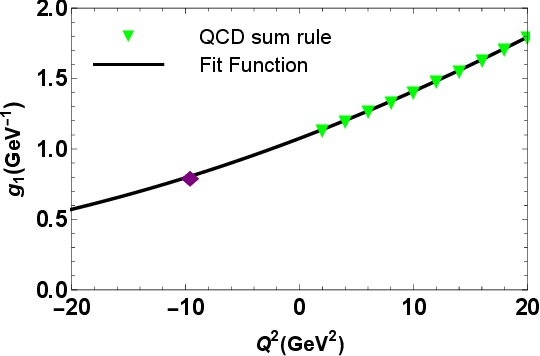}
\includegraphics[totalheight=6cm,width=8cm]{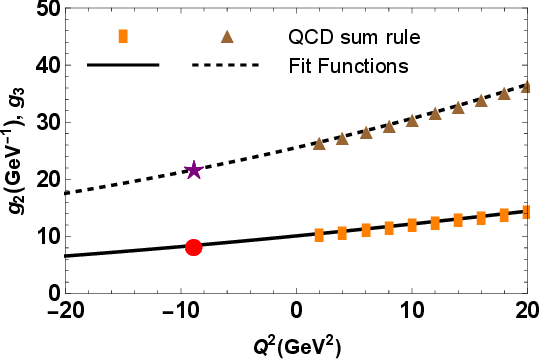}
\end{center}
\caption{Left panel: QCD data and fit function $%
\mathcal{Z}_1(Q^{2})$ for the strong coupling $g_1$. The diamond marks
the point $\mathcal{Z}_1(-m_{J\protect%
\psi}^{2})$. Right panel:
SR data and
extrapolating functions $\mathcal{Z}_2 (Q^{2})$ (solid) and $%
\mathcal{Z}_3
(Q^{2})$ (dashed). The star and circle on the fit
functions are at the point $Q^{2}=-m_{\eta_{c}}^2$. }
\label{fig:Fit}
\end{figure}

\end{widetext}

The sum rule for the form factor $g_{1}(q^{2})$ is applicable in the region $%
q^{2}<0$. But $g_{1}(q^{2})$ determines the coupling $g_{1}$ at the mass
shell $q^{2}=m_{J/\psi }^{2}$. For that reason, we employ the function $%
g_{1}(Q^{2})$ where $Q^{2}=-q^{2}$ and apply it in our studies. The SR
predictions for $g_{1}(Q^{2})$ are plotted in Fig.\ \ref{fig:Fit}, where $%
Q^{2}$ changes in the interval $Q^{2}=2-20~\mathrm{GeV}^{2}$.

It has been noted above that $g_{1}$ should be extracted at $q^{2}=m_{J/\psi
}^{2}$, i.e., at $Q^{2}=-m_{J/\psi }^{2}$. But at that point one cannot use
directly the SR method. To avoid this problem, we employ the function $%
\mathcal{Z}_{1}(Q^{2})$ which at $Q^{2}>0$ amounts to the SR data $%
g_{1}(Q^{2})$, but can be extrapolated to the region $Q^{2}<0$. For these
purposes, we employ 
\begin{equation}
\mathcal{Z}_{i}(Q^{2})=\mathcal{Z}_{i}^{0}\mathrm{\exp }\left[ z_{i}^{1}%
\frac{Q^{2}}{m^{2}}+z_{i}^{2}\left( \frac{Q^{2}}{m^{2}}\right) ^{2}\right] ,
\label{eq:FitF}
\end{equation}%
where $\mathcal{Z}_{i}^{0}$, $z_{i}^{1}$, and $z_{i}^{2}$ are fitted
constants. From confronting of the SR data and Eq.\ (\ref{eq:FitF}), it is
easy to fix 
\begin{equation}
\mathcal{Z}_{1}^{0}=1.076~\mathrm{GeV}^{-1},\ z_{1}^{1}=1.249,\text{ }%
z_{1}^{2}=-0.286.  \label{eq:FF1}
\end{equation}%
The function $\mathcal{Z}_{1}(Q^{2})$ is drawn in Fig.\ \ref{fig:Fit}, where
agreement with the sum rule's data is evident. For $g_{1}$, one gets 
\begin{equation}
g_{1}\equiv \mathcal{Z}_{1}(-m_{J/\psi }^{2})=(8.06\pm 1.47)\times 10^{-1}\ 
\mathrm{GeV}^{-1}.  \label{eq:g1}
\end{equation}

Partial width of the decay $X\rightarrow J/\psi J/\psi $ is given by the
formula%
\begin{equation}
\Gamma \left[ X\rightarrow J/\psi J/\psi \right] =\frac{g_{1}^{2}\lambda _{1}%
}{2\cdot 80\pi m^{2}}(m^{4}-3m^{2}m_{J/\psi }^{2}+6m_{J/\psi }^{4}).
\label{eq:PDw2}
\end{equation}%
In Eq.\ (\ref{eq:PDw2}), $\lambda _{1}=\lambda (m,m_{J/\psi },m_{J/\psi })$
is defined as 
\begin{equation}
\lambda (x,y,z)=\frac{\sqrt{%
x^{4}+y^{4}+z^{4}-2(x^{2}y^{2}+x^{2}z^{2}+y^{2}z^{2})}}{2x}.
\end{equation}%
Finally, one gets 
\begin{equation}
\Gamma \left[ X\rightarrow J/\psi J/\psi \right] =(41.1\pm 10.9)~\mathrm{MeV}%
.  \label{eq:DW2}
\end{equation}%
The errors above arise owing to ambiguities of $g_{1}$ and masses of
particles $J/\psi $ and $X$ in Eq.\ (\ref{eq:PDw2}).

\subsection{$X\rightarrow \protect\eta _{c}\protect\eta _{c}$}


Partial width of this process is determined by the coupling $g_{2}$ at the
vertex $X\eta _{c}\eta _{c}$. To obtain corresponding form factor $%
g_{2}(q^{2})$, one has to consider the correlator%
\begin{equation}
\Pi _{\mu \nu }(p,p^{\prime })=i^{2}\int d^{4}xd^{4}ye^{ip^{\prime
}y}e^{-ipx}\langle 0|\mathcal{T}\{I^{\eta _{c}}(y)I^{\eta _{c}}(0)I_{\mu \nu
}^{\dagger }(x)\}|0\rangle .  \label{eq:CF7}
\end{equation}%
The interpolating currents of the meson $\eta _{c}$ is 
\begin{equation}
\ I^{\eta _{b}}(x)=\overline{c}_{i}(x)i\gamma _{5}c_{i}(x)  \label{eq:C3}
\end{equation}

To calculate $\Pi _{\mu \nu }^{\mathrm{Phys}}(p,p^{\prime })$ we use the
matrix element 
\begin{equation}
\langle 0|I^{\eta _{c}}|\eta _{c}(p^{\prime })\rangle =\frac{f_{\eta
_{c}}m_{\eta _{c}}^{2}}{2m_{c}},  \label{eq:ME4}
\end{equation}%
where $f_{\eta _{c}}$, and $m_{\eta _{c}}$ are the decay constant and mass
of the meson $\eta _{c}$. The vertex $X\eta _{c}\eta _{c}$ is determined by
the expression \cite{Agaev:2024pil}%
\begin{equation}
\langle \eta _{c}(p^{\prime })\eta _{c}(q)|X(p,\epsilon )\rangle
=g_{2}(q^{2})\epsilon _{\alpha \beta }(p)p^{\prime \alpha }p^{\prime \beta }.
\end{equation}%
Then $\Pi _{\mu \nu }^{\mathrm{Phys}}(p,p^{\prime })$ is 
\begin{eqnarray}
&&\Pi _{\mu \nu }^{\mathrm{Phys}}(p,p^{\prime })=g_{2}(q^{2})\frac{\Lambda
f_{\eta _{c}}^{2}m_{\eta _{c}}^{4}}{4m_{c}^{2}\left( p^{2}-m^{2}\right)
(p^{\prime 2}-m_{\eta _{c}}^{2})}\frac{1}{(q^{2}-m_{\eta _{c}}^{2})}\left\{ 
\frac{1}{12m^{2}}\left[ m^{4}-2m^{2}(m_{\eta _{c}}^{2}+q^{2})\right. \right.
\notag \\
&&\left. \left. +(m_{\eta _{c}}^{2}-q^{2})^{2}\right] g_{\mu \nu }+p_{\mu
}^{\prime }p_{\nu }^{\prime }+\text{other terms}\right\} .  \label{eq:CF4}
\end{eqnarray}%
The function $\Pi _{\mu \nu }^{\mathrm{OPE}}(p,p^{\prime })$ reads 
\begin{eqnarray}
&&\Pi _{\mu \nu }^{\mathrm{OPE}}(p,p^{\prime })=2\int
d^{4}xd^{4}ye^{ip^{\prime }y}e^{-ipx}\left\{ \mathrm{Tr}\left[ \gamma
_{5}S_{c}^{ia}(y-x)\gamma _{\mu }\widetilde{S}_{c}^{jb}(-x)\gamma _{5}%
\widetilde{S}_{c}^{aj}(x)\gamma _{\nu }S_{c}^{bi}(x-y)\right] \right.  \notag
\\
&&\left. -\mathrm{Tr}\left[ \gamma _{5}S_{c}^{ia}(y-x)\gamma _{\mu }%
\widetilde{S}_{c}^{jb}(-x)\gamma _{5}\widetilde{S}_{c}^{bj}(x)\gamma _{\nu
}S_{c}^{ai}(x-y)\right] \right\} .  \label{eq:CF5}
\end{eqnarray}%
The terms in $\Pi _{\mu \nu }^{\mathrm{Phys}}(p,p^{\prime })$ and $\Pi _{\mu
\nu }^{\mathrm{OPE}}(p,p^{\prime })$ have the identical Lorentz structures.
The sum rule for $g_{2}(q^{2})$ is found using terms $\sim g_{\mu \nu }$ and
corresponding amplitudes $\Pi _{2}^{\mathrm{Phys}}(p^{2},p^{\prime 2},q^{2})$
and $\Pi _{2}^{\mathrm{OPE}}(p^{2},p^{\prime 2},q^{2})$. As a result, we get

\begin{equation}
g_{2}(q^{2})=\frac{48m^{2}m_{c}^{2}(q^{2}-m_{\eta _{c}}^{2})}{\Lambda
f_{\eta _{c}}^{2}m_{\eta _{c}}^{4}[m^{4}-2m^{2}(m_{\eta
_{c}}^{2}+q^{2})+(m_{\eta _{c}}^{2}-q^{2})^{2}]}e^{m^{2}/M_{1}^{2}}e^{m_{%
\eta _{c}}^{2}/M_{2}^{2}}\Pi _{2}(\mathbf{M}^{2},\mathbf{s}_{0},q^{2}),
\label{eq:SRCoup}
\end{equation}%
where $\Pi _{2}(\mathbf{M}^{2},\mathbf{s}_{0},q^{2})$ is the amplitude $\Pi
_{2}^{\mathrm{OPE}}(p^{2},p^{\prime 2},q^{2})$ after relevant
transformations and subtractions.

Operations to extract $g_{2}(q^{2})$ have been explained above therefore we
omit further details. In computations, we use the mass $m_{\eta
_{c}}=(2984.1\pm 0.4)\ \mathrm{MeV}$ \cite{PDG:2024} and decay constant $%
f_{\eta _{c}}=(398.1\pm 1.0)\ \mathrm{MeV}$ of the pseudoscalar charmonium $%
\eta _{c}$ \cite{Hatton:2020qhk}. In the $\eta _{c}$ meson channel we employ
the regions 
\begin{equation}
M_{2}^{2}\in \lbrack 3.5,4.5]~\mathrm{GeV}^{2},\ s_{0}^{\prime }\in \lbrack
11,12]~\mathrm{GeV}^{2}.
\end{equation}

The extrapolating function $\mathcal{Z}_{2}(Q^{2})$ with parameters $%
\mathcal{Z}_{2}^{0}=10.094~\mathrm{GeV}^{-1}$, $z_{2}^{1}=0.859$, and $%
z_{2}^{2}=0.170$ nicely agrees with QCD predictions (see, Fig.\ \ref{fig:Fit}%
). The strong coupling $g_{2}$ amounts to 
\begin{equation}
g_{2}\equiv \mathcal{Z}_{2}(-m_{\eta _{c}}^{2})=(8.41\pm 1.52)\ \mathrm{GeV}%
^{-1}.
\end{equation}%
The partial width of this decay is equal to 
\begin{equation}
\Gamma \left[ X\rightarrow \eta _{c}\eta _{c}\right] =g_{2}^{2}\frac{\lambda
_{2}}{2\cdot 960\pi m^{2}}(m^{2}-4m_{\eta _{c}}^{2})^{2},
\end{equation}%
where $\lambda _{2}=\lambda (m,m_{\eta _{c}},m_{\eta _{c}})$.

The process $X\rightarrow \eta _{c}\eta _{c}$ has the width 
\begin{equation}
\Gamma \left[ X\rightarrow \eta _{c}\eta _{c}\right] =(24.7\pm 7.7)~\mathrm{%
MeV}.  \label{eq:DW20}
\end{equation}


\subsection{$X\rightarrow \protect\chi _{c1}(1P)\protect\eta _{c}$}


The partial width of this decay is found using the strong coupling $g_{3}$
at the vertex $X\chi _{c1}\eta _{c}$. This coupling is obtained by exploring
the correlation function 
\begin{equation}
\Pi _{\mu \alpha \beta }(p,p^{\prime })=i^{2}\int d^{4}xd^{4}ye^{ip^{\prime
}y}e^{-ipx}\langle 0|\mathcal{T}\{I_{\mu }^{\chi _{c1}}(y)I^{\eta
_{c}}(0)I_{\alpha \beta }^{\dagger }(x)\}|0\rangle ,
\end{equation}%
where $I_{\mu }^{\chi _{c1}}(x)$ is the interpolating current for the meson $%
\chi _{c1}$%
\begin{equation}
I_{\mu }^{\chi _{c1}}(x)=\overline{c}_{i}(x)\gamma _{5}\gamma _{\mu
}c_{i}(x).
\end{equation}

Calculation of $\Pi _{\mu \alpha \beta }^{\mathrm{Phys}}(p,p^{\prime })$
does not differ significantly from analysis of other correlation functions
and leads to the expression 
\begin{eqnarray}
&&\Pi _{\mu \alpha \beta }^{\mathrm{Phys}}(p,p^{\prime })=g_{3}(q^{2})\frac{%
\Lambda f_{\eta _{c}}m_{\eta _{c}}^{2}f_{\chi _{c1}}m_{\chi _{c1}}}{%
4m_{c}\left( p^{2}-m^{2}\right) (p^{\prime 2}-m_{\chi _{c1}}^{2})}\frac{1}{%
(q^{2}-m_{\eta _{c}}^{2})}\left\{ g_{\alpha \beta }p_{\mu }^{\prime }-\frac{%
(m^{2}+m_{\chi _{c1}}^{2}-q^{2})^{2}}{6m^{2}m_{\chi _{c1}}^{2}}g_{\mu \alpha
}p_{\beta }^{\prime }\right.  \notag \\
&&\left. +\text{ other contributions}\right\} .  \label{eq:CF8}
\end{eqnarray}%
This correlator has been calculated by employing the matrix elements of the
mesons $\chi _{c1}$, $\eta _{c}$ and vertex $\langle \chi _{c1}(p^{\prime
},\varepsilon )\eta _{c}(q)|X(p,\epsilon )\rangle $. The matrix element of
the charmonium $\eta _{c}$ has been introduced in the previous subsection.
For the remaining two elements, we have used%
\begin{equation}
\langle 0|I_{\mu }^{\chi _{c1}}|\chi _{c1}(p^{\prime },\varepsilon )\rangle
=f_{\chi _{c1}}m_{\chi _{c1}}\varepsilon _{\mu }(p^{\prime }),
\end{equation}%
and 
\begin{equation}
\langle \chi _{c1}(p^{\prime },\varepsilon )\eta _{c}(q)|X(p,\epsilon
)\rangle =g_{3}(q^{2})\epsilon _{\rho \sigma }(p)q^{\rho }\varepsilon
^{\sigma }(p^{\prime }).
\end{equation}%
Above $m_{\chi _{c1}}=(3510.67\pm 0.05)\ \mathrm{MeV}$ and $f_{\chi
_{c1}}=149\ \mathrm{MeV}$ are the mass and decay constant of the meson $\chi
_{c1}$ \cite{PDG:2024,Lakhina:2006vg}, respectively.

The correlation function $\Pi _{\mu \alpha \beta }^{\mathrm{OPE}%
}(p,p^{\prime })$ is given by the expression 
\begin{eqnarray}
&&\Pi _{\mu \alpha \beta }^{\mathrm{OPE}}(p,p^{\prime })=2\int
d^{4}xd^{4}ye^{ip^{\prime }y}e^{-ipx}\left\{ \mathrm{Tr}\left[ \gamma _{\mu
}\gamma _{5}S_{c}^{ia}(y-x)\gamma _{\alpha }\widetilde{S}_{c}^{jb}(-x)\gamma
_{5}\widetilde{S}_{c}^{aj}(x)\gamma _{\beta }S_{c}^{bi}(x-y)\right] \right. 
\notag \\
&&\left. -\mathrm{Tr}\left[ \gamma _{\mu }\gamma _{5}S_{c}^{ia}(y-x)\gamma
_{\alpha }\widetilde{S}_{c}^{jb}(-x)\gamma _{5}\widetilde{S}%
_{c}^{bj}(x)\gamma _{\beta }S_{c}^{ai}(x-y)\right] \right\} .
\end{eqnarray}%
Having used the amplitudes which correspond to structures $g_{\alpha \beta
}p_{\mu }^{\prime }$, we find the SR for the form factor $g_{3}(q^{2})$ 
\begin{equation}
g_{3}(q^{2})=\frac{4m_{c}(q^{2}-m_{\eta _{c}}^{2})}{\Lambda f_{\eta
_{c}}m_{\eta _{c}}^{2}f_{\chi _{c1}}m_{\chi _{c1}}}e^{m^{2}/M_{1}^{2}}e^{m_{%
\chi _{c1}}/M_{2}^{2}}\Pi _{3}(\mathbf{M}^{2},\mathbf{s}_{0},q^{2}).
\end{equation}%
We use for parameters $M_{2}^{2}$ and $s_{0}^{\prime }$ in the $\chi _{c1}$
channel the regions 
\begin{equation}
M_{2}^{2}\in \lbrack 4,5]~\mathrm{GeV}^{2},\ s_{0}^{\prime }\in \lbrack
13,15]~\mathrm{GeV}^{2}.
\end{equation}%
The SR data are extracted for $Q^{2}=2-20~\mathrm{GeV}^{2}$ and demonstrated
in Fig.\ \ref{fig:Fit}. The corresponding fit function has parameters $%
\mathcal{Z}_{3}^{0}=25.562$, $z_{3}^{1}=0.803$, and $z_{3}^{2}=-0.040$. For $%
g_{3}$, we find 
\begin{equation}
g_{3}\equiv \mathcal{Z}_{3}(-m_{\eta _{c}}^{2})=21.66\pm 3.90.
\end{equation}%
The width of the decay $X\rightarrow \chi _{c1}\eta _{c}$ is calculated by
means of the formula%
\begin{equation}
\Gamma \left[ X\rightarrow \chi _{c1}\eta _{c}\right] =g_{3}^{2}\frac{%
\lambda _{3}}{40\pi m^{2}}|M|^{2},
\end{equation}%
where 
\begin{eqnarray}
&&|M|^{2}=\frac{1}{24m^{4}m_{\chi _{c1}}^{2}}\left[ m^{8}-2m^{2}(2m_{\eta
_{c}}^{2}-3m_{\chi _{c1}}^{2})(m_{\chi _{c1}}^{2}-m_{\eta
_{c}}^{2})^{2}+(m_{\chi _{c1}}^{2}-m_{\eta _{c}}^{2})^{4}\right.  \notag \\
&&\left. +m^{6}\left( 6m_{\chi _{c1}}^{2}-4m_{\eta _{c}}^{2}\right)
+2m^{4}(3m_{\eta _{c}}^{4}-8m_{\chi _{c1}}^{2}m_{\eta _{c}}^{2}-7m_{\chi
_{c1}}^{4})\right]
\end{eqnarray}%
and $\lambda _{3}=\lambda (m,m_{\chi _{c1}},m_{\eta _{c}}).$

For the width of this process, we get 
\begin{equation}
\Gamma \left[ X\rightarrow \chi _{c1}\eta _{c}\right] =(32.5\pm 15.7)~%
\mathrm{MeV}.
\end{equation}


\section{Subleading channels of $X$}

\label{sec:Widths2}


The tetraquark $X$ transforms to ordinary mesons after annihilation of $c%
\overline{c}$ quarks to light quark-antiquark pairs \cite%
{Becchi:2020mjz,Becchi:2020uvq,Agaev:2023ara} and generation of $DD$ mesons
with appropriate quantum numbers. Here, we consider the processes $%
X\rightarrow D^{+}D^{-}$,$\ D^{0}\overline{D}^{0}$, $D^{\ast +}D^{\ast -}$,$%
\ D^{\ast 0}\overline{D}^{\ast 0}$, $D_{s}^{+}D_{s}^{-}$ and $D_{s}^{\ast
+}D_{s}^{\ast -}$.

Note that the correlation functions of decays, for instance, $X\rightarrow
D^{+}D^{-}$ and $X\rightarrow D^{0}\overline{D}^{0}$ differ from each other
only by propagators of $d$ and $u$ quarks. Because we use the approximation $%
m_{u}=m_{d}=0$, and also neglect small numerical differences in the masses
of charged and neutral $D$ mesons, the decays $X\rightarrow D^{+}D^{-}$ and $%
X\rightarrow D^{0}\overline{D}^{0}$ have the same widths.


\subsection{Processes $X\rightarrow D^{\ast +}D^{\ast -}$ and$\ D^{\ast 0}%
\overline{D}^{\ast 0}$}


Here, we concentrate on the decay $X\rightarrow D^{\ast +}D^{\ast -}$. To
determine the strong coupling $g_{4}$ at the tetraquark-meson-meson vertex $%
XD^{\ast +}D^{\ast -}$, we study the correlation function%
\begin{equation}
\widehat{\Pi }_{\mu \nu \alpha \beta }(p,p^{\prime })=i^{2}\int
d^{4}xd^{4}ye^{ip^{\prime }y}e^{-ipx}\langle 0|\mathcal{T}\{I_{\mu
}^{D^{\ast +}}(y)I_{\nu }^{D^{\ast -}}(0)I_{\alpha \beta }^{\dagger
}(x)\}|0\rangle ,  \label{eq:CF1A}
\end{equation}%
where $I_{\mu }^{D^{\ast +}}(x)$ and $I_{\nu }^{D^{\ast -}}(x)$ are the
interpolating currents of the mesons $D^{\ast +}$ and $D^{\ast -}$, 
\begin{equation}
I_{\mu }^{D^{\ast +}}(x)=\overline{d}_{i}(x)\gamma _{\mu }c_{i}(x),\ I_{\nu
}^{D^{\ast -}}(x)=\overline{c}_{j}(x)\gamma _{\nu }d_{j}(x).  \label{eq:CRB}
\end{equation}

In terms of the matrix elements of the particles $X$, $D^{\ast +}$, and $%
D^{\ast -}$ the correlator $\widehat{\Pi }_{\mu \nu \alpha \beta
}(p,p^{\prime })$ is 
\begin{eqnarray}
&&\widehat{\Pi }_{\mu \nu \alpha \beta }^{\mathrm{Phys}}(p,p^{\prime })=%
\frac{\langle 0|I_{\mu }^{D^{\ast +}}|D^{\ast +}(p^{\prime },\varepsilon
_{1})\rangle }{p^{\prime 2}-m_{D^{\ast }}^{2}}\frac{\langle 0|I_{\nu
}^{D^{\ast -}}|D^{\ast -}(q,\varepsilon _{2})\rangle }{q^{2}-m_{D^{\ast
}}^{2}}\langle D^{\ast +}(p^{\prime },\varepsilon _{1})D^{\ast
-}(q,\varepsilon _{2})|X(p,\epsilon )\rangle  \notag \\
&&\times \frac{\langle X(p,\epsilon )|I_{\alpha \beta }^{\dagger }|0\rangle 
}{p^{2}-m^{2}}+\cdots ,
\end{eqnarray}%
where $m_{D^{\ast }}=(2010.26\pm 0.05)~\mathrm{MeV}$ is the mass of the
mesons $D^{\ast \pm }$, and $\varepsilon _{1\mu }$ and $\varepsilon _{2\nu }$
are their polarization vectors, respectively.

The function $\widehat{\Pi }_{\mu \nu \alpha \beta }^{\mathrm{Phys}%
}(p,p^{\prime })$ is found by means of the matrix elements 
\begin{equation}
\langle 0|I_{\mu }^{D^{\ast +}}|D^{\ast +}(p^{\prime },\varepsilon
_{1})\rangle =f_{D^{\ast }}m_{D^{\ast }}\varepsilon _{1\mu }(p^{\prime }),\
\langle 0|I_{\nu }^{D^{\ast -}}|D^{\ast -}(q,\varepsilon _{2})\rangle
=f_{D^{\ast }}m_{D^{\ast }}\varepsilon _{2\nu }(q),  \label{eq:ME2B}
\end{equation}%
with $f_{D^{\ast }}=(252.2\pm 22.66)~\mathrm{MeV}$ being the decay constants
of the mesons $D^{\ast \pm }$ \cite{Lucha:2014spa}. The vertex $\langle
D^{\ast +}(p^{\prime },\varepsilon _{1})D^{\ast -}(q,\varepsilon
_{2})|X(p,\epsilon )\rangle $ is given by Eq.\ (\ref{eq:TVV}).

A sum rule for the form factor $g_{4}(q^{2})$ is obtained by employing the
amplitude $\Pi _{4}^{\mathrm{Phys}}(p^{2},p^{\prime 2},q^{2})$ that
corresponds in $\widehat{\Pi }_{\mu \nu \alpha \beta }^{\mathrm{Phys}%
}(p,p^{\prime })$ to the term $\sim p_{\beta }p_{\nu }p_{\mu }^{\prime
}p_{\alpha }^{\prime }$. The correlator $\widehat{\Pi }_{\mu \nu \alpha
\beta }(p,p^{\prime })$ computed using the quark propagators equals to 
\begin{equation}
\widehat{\Pi }_{\mu \nu \alpha \beta }^{\mathrm{OPE}}(p,p^{\prime })=\frac{4%
}{3}\int d^{4}xd^{4}ye^{ip^{\prime }y}e^{-ipx}\langle \overline{c}c\rangle 
\mathrm{Tr}\left[ \gamma _{\mu }S_{d}^{ij}(y)\gamma _{\nu
}S_{c}^{ja}(-x){}\gamma _{\alpha }\gamma _{\beta }S_{c}^{ai}(x-y)\right] ,
\label{eq:QCDsideA}
\end{equation}%
where $S_{d}(x)$ is the $d$ quark's propagator \cite{Agaev:2020zad} and $%
\langle \overline{c}c\rangle $ vacuum matrix element of $\overline{c}c$. We
label by $\Pi _{4}^{\mathrm{OPE}}(p^{2},p^{\prime 2},q^{2})$ the amplitude
that corresponds in $\widehat{\Pi }_{\mu \nu \alpha \beta }^{\mathrm{OPE}%
}(p,p^{\prime })$ to the same structure $\sim p_{\beta }p_{\nu }p_{\mu
}^{\prime }p_{\alpha }^{\prime }$.

In what follows, we utilize the relation 
\begin{equation}
\langle \overline{c}c\rangle \approx -\frac{1}{12m_{c}}\langle \frac{\alpha
_{s}G^{2}}{\pi }\rangle  \label{eq:Conden}
\end{equation}%
between the condensates obtained in Ref.\ \cite{Shifman:1978bx}.

The sum rule for the form factor $g_{4}(q^{2})$ reads%
\begin{equation}
g_{4}(q^{2})=\frac{2m^{2}(q^{2}-m_{D^{\ast }}^{2})}{\Lambda f_{D^{\ast
}}^{2}m_{D^{\ast }}^{2}}e^{m^{2}/M_{1}^{2}}e^{m_{D^{\ast
}}^{2}/M_{2}^{2}}\Pi _{4}(\mathbf{M}^{2},\mathbf{s}_{0},q^{2}).
\label{eq:g4}
\end{equation}%
Restrictions imposed on the Borel and continuum subtraction parameters $%
\mathbf{M}^{2}=(M_{1}^{2},M_{2}^{2})$ and $\mathbf{s}_{0}=(s_{0},s_{0}^{%
\prime })$ do not differ from ones accepted and explained in previous
sections. Thus, in the tetraquark channel the pair $(M_{1}^{2},s_{0})$ is
chosen as in Eq.\ (\ref{eq:Wind1}). This allows us to avoid additional
uncertainties in quantites $m$ and $\Lambda $ in Eq.\ (\ref{eq:g4}) and
ensure fulfillment of standard constraints of SR analysis. For the $D^{\ast
-}$ meson's channel, we employ the parameters%
\begin{equation}
M_{2}^{2}\in \lbrack 3,5]~\mathrm{GeV}^{2},\ s_{0}^{\prime }\in \lbrack 6,8]~%
\mathrm{GeV}^{2},  \label{eq:Wind2}
\end{equation}%
where the border $s_{0}^{\prime }=6~\mathrm{GeV}^{2}$ is less than $%
m_{D^{\ast }(2S)}^{2}\approx 7~\mathrm{GeV}^{2}$, where $m_{D^{\ast }(2S)}$
is the mass of the first radial excitation of the vector meson $D^{\ast \pm }
$ \cite{PDG:2024}. In other words, $s_{0}^{\prime }$ effectively separates
the ground-level meson $D^{\ast \pm }$ from excited and continuum states in
the corresponding channel.

\begin{widetext}

\begin{figure}[h!]
\begin{center}
\includegraphics[totalheight=6cm,width=8cm]{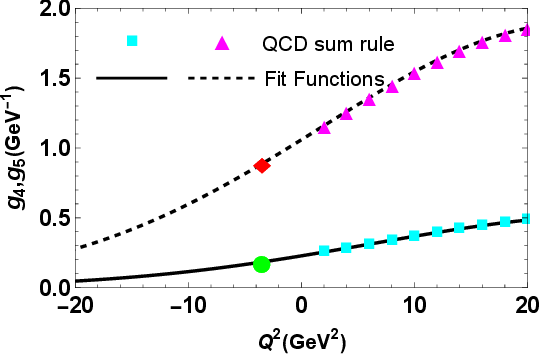}
\includegraphics[totalheight=6cm,width=8cm]{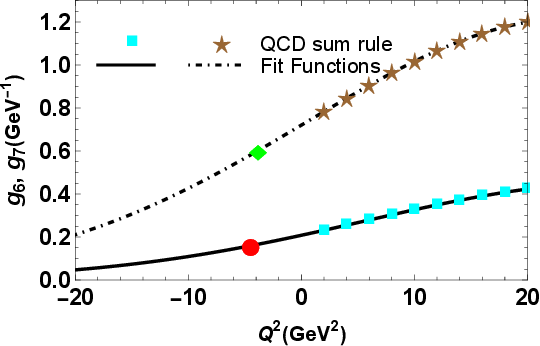}
\end{center}
\caption{Left panel: QCD data and fit functions for strong couplings $%
\mathcal{Z}_4(Q^{2})$
(solid line) and $\mathcal{Z}_5(Q^{2})$ (dashed
line). The diamond and circle mark
points $\mathcal{Z}_4(-m_{D^{*}}^{2})$ and $\mathcal{Z}_5(-m_{D}^{2})$,
respectively. Right panel:
SR data and extrapolating functions $\mathcal{Z}_6 (Q^{2})$ (solid)
and $\mathcal{Z}_7 (Q^{2})$ (dot-dashed). The diamond and circle denote the fit
functions at the points $Q^{2}=-m_{D_{s}}^{2}$ and $Q^{2}=-m_{D_{s}^{\ast
}}^2$. }
\label{fig:Fit1}
\end{figure}

\end{widetext}

To evaluate the coupling $g_{4}$ we use the SR data for $Q^{2}=2-20~\mathrm{%
GeV}^{2}$ and extrapolating function with parameters $\mathcal{Z}%
_{4}^{0}=0.228~\mathrm{GeV}^{-1}$, $z_{4}^{1}=2.537$, and $z_{4}^{2}=-1.948$%
. This function is shown in Fig.\ \ref{fig:Fit1} as solid line. The coupling 
$g_{4}$ is calculated at $q^{2}=m_{D^{\ast }}^{2}$ and is equal to 
\begin{equation}
g_{4}\equiv \mathcal{Z}_{4}(-m_{D^{\ast }}^{2})=(1.78\pm 0.34)\times
10^{-1}\ \mathrm{GeV}^{-1}.  \label{eq:G1}
\end{equation}%
The width of the decay $X\rightarrow D^{\ast +}D^{\ast -}$ is 
\begin{equation}
\Gamma \left[ X\rightarrow D^{\ast +}D^{\ast -}\right] =(11.5\pm 3.2)~%
\mathrm{MeV}.
\end{equation}

The difference between the decays $X\rightarrow D^{\ast +}D^{\ast -}$ and $%
X\rightarrow D^{\ast 0}\overline{D}^{\ast 0}$ is encoded in the masses of
the final-state mesons. With nice accuracy we adopt $\Gamma \left[
X\rightarrow D^{\ast +}D^{\ast -}\right] \approx \Gamma \left[ X\rightarrow
D^{\ast 0}\overline{D}^{\ast 0}\right] $.

\subsection{Decays $X\rightarrow D^{+}D^{-}$ and $\ D^{0}\overline{D}^{0}$}


The decay mode $X\rightarrow D^{+}D^{-}$ can be explored by means of the
correlation function 
\begin{equation}
\widehat{\Pi }_{\mu \nu }(p,p^{\prime })=i^{2}\int d^{4}xd^{4}ye^{ip^{\prime
}y}e^{-ipx}\langle 0|\mathcal{T}\{I^{D^{+}}(y)I^{D^{-}}(0)I_{\mu \nu
}^{\dagger }(x)\}|0\rangle ,
\end{equation}%
where the currents $I^{D^{+}}(x)$ and $I^{D^{-}}(x)$ are given by the
expressions%
\begin{equation}
I^{D^{+}}(x)=\overline{d}_{i}(x)i\gamma _{5}c_{i}(x),\text{ }I^{D^{-}}(x)=%
\overline{c}_{j}(x)i\gamma _{5}d_{j}(x).
\end{equation}%
The SR for the form factor $g_{5}(q^{2})$ that describes the strong
interaction of particles at the vertex $XD^{+}D^{-}$ is derived by computing the
correlators $\widehat{\Pi }_{\mu \nu }^{\mathrm{Phys}}(p,p^{\prime })$ and $%
\widehat{\Pi }_{\mu \nu }^{\mathrm{OPE}}(p,p^{\prime })$ and equating them
to get SR identity.

We find $\widehat{\Pi }_{\mu \nu }^{\mathrm{Phys}}(p,p^{\prime })$ by means
of matrix elements 
\begin{equation}
\langle 0|I^{D^{\pm }}|D^{\pm }\rangle =\frac{f_{D}m_{D}^{2}}{m_{c}},
\end{equation}%
and 
\begin{equation}
\langle D^{+}(p^{\prime })D^{-}(q)|X(p,\epsilon )\rangle
=g_{5}(q^{2})\epsilon _{\alpha \beta }(p)p^{\prime \alpha }p^{\prime \beta },
\end{equation}%
where $m_{D}=(1869.66\pm 0.05)~\mathrm{MeV}$ and $f_{D}=(211.9\pm 1.1)~%
\mathrm{MeV}$ are the mass and decay constant of mesons $D^{\pm }$ \cite%
{PDG:2024,Rosner:2015wva}, respectively. After some manipulations, one gets 
\begin{eqnarray}
&&\widehat{\Pi }_{\mu \nu }^{\mathrm{Phys}}(p,p^{\prime })=\frac{%
g_{5}(q^{2})\Lambda f_{D}^{2}m_{D}^{4}}{m_{c}^{2}\left( p^{2}-m^{2}\right)
\left( p^{\prime 2}-m_{D}^{2}\right) \left( q^{2}-m_{D}^{2}\right) }\left[ 
\frac{m^{4}-2m^{2}(m_{D}^{2}+q^{2})+(m_{D}^{2}-q^{2})^{2}}{12m^{2}}g_{\mu
\nu }\right.  \notag \\
&&\left. +p_{\mu }^{\prime }p_{\nu }^{\prime }+\text{other contributions}%
\right] .
\end{eqnarray}%
For $\widehat{\Pi }_{\mu \nu }^{\mathrm{OPE}}(p,p^{\prime })$, we find%
\begin{equation}
\widehat{\Pi }_{\mu \nu }^{\mathrm{OPE}}(p,p^{\prime })=-\frac{4}{3}\int
d^{4}xd^{4}ye^{ip^{\prime }y}e^{-ipx}\langle \overline{c}c\rangle \mathrm{Tr}%
\left[ \gamma _{5}{}S_{d}^{ij}(y)\gamma _{5}S_{c}^{ja}(-x)\gamma _{\mu
}\gamma _{\nu }S_{c}^{ai}(x-y)\right] .  \label{eq:QCDsideB}
\end{equation}%
To obtain the sum rule for $g_{5}(q^{2})$, we employ the amplitudes $\Pi
_{5}^{\mathrm{Phys}}(p^{2},p^{\prime 2},q^{2})$ and $\Pi _{5}^{\mathrm{OPE}%
}(p^{2},p^{\prime 2},q^{2})$ corresponding to structures $g_{\mu \nu }$ and
obtain 
\begin{equation}
g_{5}(q^{2})=\frac{12m^{2}m_{c}^{2}(q^{2}-m_{D}^{2})}{\Lambda
f_{D}^{2}m_{D}^{4}[m^{4}-2m^{2}(m_{D}^{2}+q^{2})+(m_{D}^{2}-q^{2})^{2}]}%
e^{m^{2}/M_{1}^{2}}e^{m_{D}^{2}/M_{2}^{2}}\Pi _{5}(\mathbf{M}^{2},\mathbf{s}%
_{0},q^{2}).
\end{equation}

In computations the following parameters have been utilized 
\begin{equation}
M_{2}^{2}\in \lbrack 2.5,3.5]~\mathrm{GeV}^{2},\ s_{0}^{\prime }\in \lbrack
4.5,5.5]~\mathrm{GeV}^{2}.
\end{equation}%
Fit function $\mathcal{Z}_{5}(Q^{2})$ with $\mathcal{Z}_{5}^{0}=1.059~%
\mathrm{GeV}^{-1}$, $z_{5}^{1}=2.074$, and $z_{5}^{2}=-1.849$ allows one to
estimate the coupling $g_{5}$ which reads 
\begin{equation}
g_{5}\equiv \mathcal{Z}_{5}(-m_{D}^{2})=(8.87\pm 1.69)\times 10^{-1}\ 
\mathrm{GeV}^{-1}.
\end{equation}%
The function $\mathcal{Z}_{5}(Q^{2})$ and relevant sum rule data are plotted
in Fig.\ \ref{fig:Fit1}.

The partial width of the channel $X\rightarrow D^{+}D^{-}$ amounts to 
\begin{equation}
\Gamma \left[ X\rightarrow D^{+}D^{-}\right] =(14.4\pm 3.9)~\mathrm{MeV}.
\end{equation}%
The width of the process $X\rightarrow D^{0}\overline{D}^{0}$, as it has
been explained above, is approximately equal to $\Gamma \left[ X\rightarrow
D^{+}D^{-}\right] $.


\subsection{ Channels $X\rightarrow $ $D_{s}^{\ast +}D_{s}^{\ast -}$ and $%
D_{s}^{+}D_{s}^{-}$}


Exploration of these processes does not differ considerably from studies of
the modes considered in the previous subsections: One has to take into
account some substitutions in the correlation functions and in parameters of
new final-state mesons. Indeed, the correlators of the channels $%
X\rightarrow $ $D_{s}^{\ast +}D_{s}^{\ast -}$ and $D_{s}^{+}D_{s}^{-}$ can
easily be obtained from Eqs.\ (\ref{eq:QCDsideA}) and (\ref{eq:QCDsideB})
after replacing $S_{d}^{ji}(y)\rightarrow S_{s}^{ji}(y)$. \ For instance,
for the decay $X\rightarrow $ $D_{s}^{\ast +}D_{s}^{\ast -}$, we have 
\begin{equation}
\widetilde{\Pi }_{\mu \nu \alpha \beta }^{\mathrm{OPE}}(p,p^{\prime })=\frac{%
4}{3}\int d^{4}xd^{4}ye^{ip^{\prime }y}e^{-ipx}\langle \overline{c}c\rangle 
\mathrm{Tr}\left[ \gamma _{\mu }S_{s}^{ij}(y)\gamma _{\nu
}S_{c}^{ja}(-x){}\gamma _{\alpha }\gamma _{\beta }S_{c}^{ai}(x-y)\right] .
\end{equation}

It is worth noting that in calculations we take into account terms $\sim
m_{s}=(93.5\pm 0.8)~\mathrm{MeV}$, which appear due to the propagator $%
S_{s}^{ji}(y)$ (in both decays), and the matrix element (for production of
the pseudoscalar mesons) 
\begin{equation}
\langle 0|I^{D_{s}^{\pm }}|D^{\pm }\rangle =\frac{f_{D_{s}}m_{D_{s}}^{2}}{%
m_{c}+m_{s}}.
\end{equation}%
But, at the same time, we neglect contributions proportional to $m_{s}^{2}$.

The parameters of the mesons $D_{s}^{(\ast )\pm }$ have the following values 
\cite{PDG:2024,Rosner:2015wva,Lubicz:2016bbi} 
\begin{eqnarray}
m_{D_{s}} &=&(1969.0\pm 1.4)~\mathrm{MeV},\ f_{D_{s}}=(249.0\pm 1.2)~\mathrm{%
MeV},  \notag \\
m_{D_{s}^{\ast }} &=&(2112.2\pm 0.4)~\mathrm{MeV},\ f_{D_{s}^{\ast
}}=(268.8\pm 6.5)~\mathrm{MeV}.
\end{eqnarray}

These decays are characterized by the strong couplings $g_{6}$ and $g_{7}$
at the vertices $XD_{s}^{\ast +}D_{s}^{\ast -}$ and $XD_{s}^{+}D_{s}^{-}$,
respectively. \ In the case of the process $X\rightarrow D_{s}^{\ast
+}D_{s}^{\ast -}$ we obtain the following prediction for $g_{6}$%
\begin{equation}
g_{6}\equiv \mathcal{Z}_{6}(-m_{D_{s}^{\ast }}^{2})=(1.6\pm 0.3)\times
10^{-1}\ \mathrm{GeV}^{-1}.
\end{equation}%
To estimate $g_{6}$, we have employed the extrapolation function $\mathcal{Z}%
_{6}(Q^{2})$ with parameters $\mathcal{Z}_{6}^{0}=0.208~\mathrm{GeV}^{-1}$, $%
z_{6}^{1}=2.391$, and $z_{6}^{2}=-1.840$. The corresponding SR data have
been found using for $M_{2}^{2}$,$\ s_{0}^{\prime }$ in $D_{s}^{\ast }$
channel the regions 
\begin{equation}
M_{2}^{2}\in \lbrack 3,6]~\mathrm{GeV}^{2},\ s_{0}^{\prime }\in \lbrack 6,8]~%
\mathrm{GeV}^{2}.
\end{equation}

The coupling $g_{7}$ amounts to 
\begin{equation}
g_{7}\equiv \mathcal{Z}_{7}(-m_{D_{s}}^{2})=(6.01\pm 1.14)\times 10^{-1}\ 
\mathrm{GeV}^{-1}.
\end{equation}%
The function $\mathcal{Z}_{7}(Q^{2})$ is determined by the constants $%
\mathcal{Z}_{7}^{0}=0.721~\mathrm{GeV}^{-1}$, $z_{7}^{1}=1.898$, and $%
z_{7}^{2}=-1.714$. The relevant SR data and extrapolating functions $%
\mathcal{Z}_{6}(Q^{2})$ and $\mathcal{Z}_{7}(Q^{2})$ are plotted in Fig.\ %
\ref{fig:Fit1} as well.

The widths of these channels are 
\begin{equation}
\Gamma \left[ X\rightarrow D_{s}^{\ast +}D_{s}^{\ast -}\right] =(8.6\pm 2.4)~%
\mathrm{MeV},
\end{equation}%
and 
\begin{equation}
\Gamma \left[ X\rightarrow D_{s}^{+}D_{s}^{-}\right] =(5.8\pm 1.6)~\mathrm{%
MeV},
\end{equation}%
respectively.

With information about partial widths of three leading and six subleading
decays at hand one can easily estimate the full decay width of the tensor
tetraquark $X$%
\begin{equation}
\Gamma \lbrack X]=(165\pm 23)~\mathrm{MeV}.
\end{equation}

\section{Discussions and final notes}

\label{sec:Conc}

In the present article, we have explored the tensor diquark-antidiquark
state $X$ \ with inner organization $C\gamma _{\mu }\otimes \gamma _{\nu
}C+C\gamma _{\nu }\otimes \gamma _{\mu }C$ by computing its mass and decay
width. The mass $m=(6609\pm 50)~\mathrm{MeV}$ of this tetraquark is
compatible with experimental data provided one includes into analysis the
theoretical ambiguities of SR method and experimental errors. In fact, mass
of $X(6600)$ extracted by CMS collaboration is $6593_{-14}^{+15}\pm 25~%
\mathrm{MeV}$ \cite{CMS:2026tiu}, whereas ATLAS group reported $%
6630_{-10}^{+80}\pm 50~\mathrm{MeV.}$

The situation with the width of the resonance $X(6600)$ is not quite clear:
Our prediction $(165\pm 23)~\mathrm{MeV}$ for $\Gamma \lbrack X]$ is smaller
than ones of the CMS and ATLAS collaborations. Thus, the width of this
resonance was estimated as 
\begin{equation}
446_{-54}^{+66}\pm 87~\mathrm{MeV},\ 350\pm 110_{-40}^{+110}~\mathrm{MeV},
\end{equation}%
respectively, which are, in central values, considerably larger than our
result. As is seen, these data suffer from large errors, and only in the
lower limit the ATLAS's datum is comparable with upper value of $\Gamma
\lbrack X]$. This problem can be solved by improving accuracy both of \ the
theoretical analysis and experimental measurements; more precise
experimental information is a key issue in the case under discussion.

The diquark-antidiquark states are tightly bound structures and, as a
result, emerge as relatively narrow resonances. The width of $X(6600)$
reported by experimental groups, if restrict ourselves by central values,
may be explained by existence in $X$ an additional components and
alternative decay channels. For example, considerable molecule-type
component may improve agreement of $\Gamma \lbrack X]$ with experimental
data because, as usual, a width of such structures are larger than ones of
diquark-antidiquark states. It is also known that apart from the strong
decays considered here, there are processes which run through alternative
mechanisms \cite{Becchi:2020uvq,Karliner:2016zzc,Esposito:2018cwh}. Thus,
fully-heavy tetraqurks owing to annihilations of heavy quark-antiquark pairs
in $X$ can transform to $2\gamma $, $J/\psi l^{+}l^{-}$ or to four leptons $%
l_{1}^{+}l_{1}^{-}l_{2}^{+}l_{2}^{-}$ . The full width of the tetraquark may
receive considerable contributions from these modes. As is seen,
comprehensive analysis of $X$ requires additional studies and efforts which
are beyond the scope of current work. At present level of our knowledge, we
consider the tensor diquark-antidiquark state $X$ as a part of the resonance 
$X(6600)$.

The SR result permits us to estimate also the mass of the first radially
excited tensor state $X(2S)$. While calculating the mass $m$ of the
ground-level particle $X=X(1S)$ we have fixed parameters $M^{2}$, and$\
s_{0} $ by applying the usual constraints of the sum rule method. The
continuum subtraction parameter $s_{0}$ has been introduced to separate the
contribution of the ground-level particle from ones of excited and continuum
states. This means that the mass $m^{\prime }$ of the excited state $X(2S)$
should be $m^{\prime }\geq \sqrt{s_{0}}$. In other words, $m^{\prime }\geq
7211\ \mathrm{MeV}$ which implies the $600\ \mathrm{MeV}$ mass gap between
them. This difference is reasonable estimate for the tetraquarks built of $4$
charm quarks, because the mass splittings between ordinary $c\overline{c}$
mesons%
\begin{equation}
m[\eta _{c}(2S)]-m[\eta _{c}(1S)]\approx 650~\mathrm{MeV,\ }m[\psi
(2S)]-m[J/\psi ]\approx 590~\mathrm{MeV,}
\end{equation}%
are of the same order.

In our previous paper \cite{Agaev:2023rpj} we interpreted the resonance $%
X(7300)$ as a radial excitation of $X(6600)$. It seems that a similar
assignment is valid for the tensor tetraquark $X(2S)$ as well, because its
mass $m^{\prime }$ is consistent with measurements. In reality, the mass of
the resonance $X(7300)$ in accordance to ATLAS collaboration is equal to 
\begin{equation}
7220\pm 30_{-40}^{+10}~\mathrm{MeV,}
\end{equation}%
whereas CMS experiment provided $7173_{-10}^{+9}\pm 13~\mathrm{MeV}$ in the
di-$J/\psi $ and $7169_{-52-72}^{+26+80}~\mathrm{MeV}$ in the $J/\psi \psi
(2S)$ channels by labeling it $X(7100)$. In both cases there are agreements
with $m^{\prime }\geq 7211~\mathrm{MeV}$.

Calculations made in the current article are useful for understanding
properties of the $X$ resonances. Bearing in mind problems with the full
width of this state and large experimental errors, we nevertheless interpret 
$X$ as an essential component of the resonance $X(6600)$. The structure $%
X(7300)/X(7100)$ may be considered as its radial excitation. The resonances $%
X(6200)$ and $X(6900)$ may be treated in diquark-antidiquark model using
other diquarks or explored in the hadronic molecule picture. The hadronic
molecules $J/\psi J/\psi $ and $J/\psi \psi (2S)$ are nice candidates to
these states. We eventually will address all these issues in our future
works.

\appendix*

\section{ The correlation function $\Pi (M^{2},s_{0})$}


This Appendix contains the explicit expression for the perturbative
component $\rho ^{\mathrm{pert.}}(s)$ of the spectral density and function $%
\Pi ^{\mathrm{Dim4}}(M^{2})$: They have been applied to compute the mass of
the tensor tetraquark $X$ .

The correlation function $\Pi (M^{2},s_{0})$ which appears in the SRs has
the following form 
\begin{equation}
\Pi (M^{2},s_{0})=\int_{16m_{c}^{2}}^{s_{0}}ds\rho ^{\mathrm{OPE}%
}(s)e^{-s/M^{2}}+\Pi (M^{2}).
\end{equation}

The components of the spectral density 
\begin{equation}
\rho ^{\mathrm{OPE}}(s)=\rho ^{\mathrm{pert.}}(s)+\rho ^{\mathrm{Dim4}}(s),
\end{equation}%
are given by the general expression%
\begin{equation}
\rho (s)=\int_{0}^{1}dx\int_{0}^{1-x}dy\int_{0}^{1-x-y}dz\rho (s,x,y,z),
\label{eq:A4}
\end{equation}%
where $x$, $y$, and $z$ are the Feynman parameters. The function $\Pi
(M^{2}) $ is also determined by an Eq.\ (\ref{eq:A4})-type formula with the
integrand $\Pi ^{\mathrm{Dim4}}(M^{2},x,y,z)$.

The perturbative function $\rho ^{\mathrm{pert.}}(s,x,y,z)$ is given by the
expression 
\begin{eqnarray}
&&\rho ^{\mathrm{pert.}}(s,x,y,z)=\frac{L^{2}[s,x,y,z]\Theta \lbrack
L[s,x,y,z]]}{256\pi ^{6}B^{2}(x+y)^{4}}%
[-12(m_{c}^{2}-CF(1+F)s)(m_{c}^{2}-(-1+A)A(-1+C)^{2}(1+F)^{2}s)  \notag \\
&&-4(Cm_{c}^{2}+BEm_{c}^{2}-A(1-2C+C^{2}+E)m_{c}^{2}-BCEFs-BCEF^{2}s+A(1+F)(-EF-4C^{2}(1+2F)
\notag \\
&&+2C^{3}(1+2F)+2C(1+(2+E)F))s+A^{2}((1-2C+C^{2}+E)m_{c}^{2}  \notag \\
&&-(1+F)(-EF-4C^{2}(1+2F)+2C^{3}(1+2F)+2C(1+(2+E)F))s))L[s,x,y,z]  \notag \\
&&-(BCE+A(4C^{2}-2C^{3}+E-2C(1+E))+A^{2}(-4C^{2}+2C^{3}-E+2C(1+E)))L^{2}[s,x,y,z]],
\end{eqnarray}%
where $\Theta (t)$ is the Unit Step function and 
\begin{eqnarray}
&&A=\frac{x}{x+y},\ B=\frac{xy+xz+yz}{(x+y)^{2}},\ C=\frac{x+y}{xy+xz+yz},\
D=y^{2}+(-1+z)z+y(-1+2z),  \notag \\
E &=&\frac{(-x-y)(yz(-1+y+z)+x^{2}(y+z)+xD)}{(xy+xz+yz)^{2}},\ F=\frac{xyz}{%
yz(-1+y+z)+x^{2}(y+z)+xD},\   \notag \\
G &=&-1+x+y+z,  \notag \\
L[s,x,y,z] &=&\frac{(yz+xy+xz)(-sxyzG+m_{c}^{2}(yz(-1+y+z)+x^{2}(y+z)+xD))}{%
(yz(-1+y+z)+x^{2}(y+z)+xD)^{2}}.
\end{eqnarray}

The function $\Pi ^{\mathrm{Dim4}}(s,x,y,z)$ is determined by the formula
\begin{eqnarray}
&&\Pi ^{\mathrm{Dim4}}(s,x,y,z)=\frac{A^{7}Fm_{c}^{6}\Theta \lbrack
L_{1}[s,x,y,z]]}{192\pi ^{4}x^{9}y^{2}z^{2}G^{2}B^{5}E^{3}K}\langle \frac{%
\alpha _{s}G^{2}}{\pi }\rangle \exp [-L_{1}/M^{2}](-A^{2}(-1+C)^{2}F(1+F)^{2}
\notag \\
&&\times
(C-F+CF)(Cz^{3}+F(-1+x^{3}+y^{3}+3y^{2}(-1+z)+3y(-1+z)^{2}+3z-3z^{2}+z^{3}+Cz^{3}+3x^{2}(-1+y+z)
\notag \\
&&+3x(-1+y+z)^{2}))+A(-1+C)^{2}(1+F)^{2}(C^{2}F(1+F)^{2}z^{3}+CF^{2}(1+F)(-1+2x^{3}+3y^{2}(-1+z)
\notag \\
&&+3y(-1+z)^{2}+3z-3z^{2}+3x^{2}(-1+y+z)+3x(-1+y+z)^{2})-G(xy(x^{3}-y^{3})z 
\notag \\
&&+F^{3}G^{2})+x(-C^{3}F^{2}(1+F)^{3}x^{2}+C^{2}(1+F)^{2}(2F^{2}x^{2}+2F^{3}x^{2}
\notag \\
&&+yzG(x^{3}+z^{3}))-yzG(-x^{3}-2Fx^{3}+F^{2}(-1+y+z)(3x^{2}  \notag \\
&&+3x(-1+y+z)+(-1+y+z)^{2}))-C(1+F)(F^{2}x^{2}+2F^{3}x^{2}+F^{4}x^{2}+2x^{3}yzG
\notag \\
&&-Fyz(-x^{4}+2x^{3}(-1+y+z)+6x^{2}(-1+y+z)^{2}+x(-4+4y^{3}+12y^{2}(-1+z) 
\notag \\
&&+12y(-1+z)^{2}+12z-12z^{2}+3z^{3})+(-1+y)(-1+y^{3}+4z-6z^{2}+3z^{3}+y^{2}(-3+4z)
\notag \\
&&+y(3-8z+6z^{2})))))).
\end{eqnarray}%
In this expression, we introduced additional notations

\begin{equation}
L_{1}[s,x,y,z]=\frac{Fm_{c}^{2}}{xyzG},\ K=\frac{xyzG(yz+xy+xz)}{F^{2}}.
\end{equation}

\qquad

\end{document}